\documentclass[preprint,showpacs,preprintnumbers,amsmath,amssymb]{revtex4}

\usepackage{graphicx}
\usepackage{dcolumn}
\usepackage{bm}
\usepackage[]{amssymb,epsfig,color}
\newcommand{\be}{\begin{equation}}
\newcommand{\ee}{\end{equation}}
\begin{document}

\title{Two-photon driven nonlinear dynamics and entanglement of an atom in a non uniform cavity}

\author{L. Chotorlishvili$^{1,3}$,  Z. Toklikishvili$^{2}$,   S. Wimberger$^{3}$, and J. Berakdar$^{1}$}
%

\affiliation{1 Institut f\"ur Physik, Martin-Luther Universit\"at
Halle-Wittenberg, Heinrich-Damerow-Str.4,
06120 Halle, Germany\\
2 Physics Department of the Tbilisi State University,
Chavchavadze av.3, 0128, Tbilisi, Georgia \\
3  Institut f\"ur Theoretische Physik, Universit\"at Heidelberg,
Philosophenweg 19, D-69120 Heidelberg,Germany}

\begin{abstract}

In this paper we  study   the dynamics in the  general case for a Tavis Cummings atom in a non-uniform cavity.
In addition to  the dynamical Stark shift, the center-of-mass motion of the atom and the recoil effect are
considered in both - the weak and the strong cavity atom coupling
regimes. It is shown that the spatial motion of the atom inside the
cavity in the resonant case leads to a transition between
topologically different solutions.  This effect is manifested
by a singularity in the inter-level transition spectrum.
In the non-resonant case, the spatial motion of the atom leads to a switching
of the spin orientation. In both effects, the key factor is the relation
between the values of the Stark shift and the cavity field coupling
constant.  We also investigate the entanglement of an atom in the cavity with the radiation field.
It is shown that the entanglement between the atom and the field, usually quantified  in terms
of purity, decreases with increasing the Stark shift.
\end{abstract}

\maketitle
\section{Introduction}
 Cavity quantum electrodynamics (CQED) is an active  field of research  focusing
on  the quantum nature
 of the interaction of atoms with photons in
high-finesse cavities \cite{Aoki,Mabuchi,Hood,Raimond}. Current
issues of interest includes entanglement and quantum correlations
\cite{Shevchenko,Maruyama,Buluta,Amico,Mintert}. The standard model of CQED
is a  two level  system  in a quantum cavity. This case has
 served as an example for the  realization of quantum programming protocols and for
quantum teleportation  \cite{Bruss}. A controlled system of several
atoms is also promising as a candidate for multiqubit entangled state.
 In the simplest situation of a single atom interacting
with few photons in the regime of strong coupling, the coherent
atom-photon  interaction overwhelms  incoherent dissipative
processes.The system can then be described  by the Jaynes Cummings
(JC) model \cite{Jaynes,Schleich} which captures the interaction
of a single cavity mode with a frequency resonant  with the
transition frequency of the two lowest energy levels of the atom.
The interactions between the atom and the radiation field  should,
however, involve not only the internal atomic transitions and
field states but also should account for the center-of-mass motion
of the atom and for recoil effects. Since the motion of the atom
in a cavity and inter-level transitions are connected to each
other, the dynamics in a non-uniform cavity becomes rather
complicated. For particular values of parameters, the
corresponding classical system manifests regular or chaotic
behavior:
 Different types of motion are found, including L\'evy flights and
chaotic walking of an atom in a cavity \cite{Prants}. Consequences
of the non-uniform cavity are also  decoherence effects,
 a decay of entanglement and of the  teleportation
fidelity \cite{Chotorlishvili}.\\
The main goal of the present  work is the study of the dynamics of
$su(1,1)$ Tavis-Cummings (TC) system in a non-uniform cavity
including the dynamical Stark shift (DSS) \cite{Joshi,Rybin}. This
model describes two-photon transitions between the ground and the
excited state via an intermediate state. The intermediate state
can be eliminated from the equations of motion on the cost of
introducing a dynamical Stark shift \cite{Joshi}. In particular,
we study the impact of DSS on the dynamics of TC atom in
non-uniform cavities. We identify  different types of dynamics and
possible mechanisms of switching between them. In the first part
we use a semiclassical approach, which is a natural approximation
for large mean photon numbers in the cavity. We investigate the
problem in both the strong and the weak atom-cavity coupling
regimes. In the second part we  go beyond the semiclassical
approximation and evaluate the influence of the cavity being non-uniform
 and also of DSS on the degree of entanglement between the atom
and the radiation field.

\section{Model}
The Hamiltonian of a single TC atom placed in an ideal cavity reads:
 \be
 \hat{H}=\frac{\hat{P}^2}{2m}+\big(\omega_{0}+\zeta\hat{a}^{+}\hat{a}\big)\hat{s}^{z}+
 \omega_{f}\hat{a}^{+}\hat{a}+g(x)\big(\hat{s}^{+}\hat{a}^{2}+\hat{s}^{-}(\hat{a}^{+})^{2}\big)\label{eq:ham}
.\ee
 Here $g(x)$ is the coupling constant between the atom and the radiation field,
 $\hat{a}$,  $\hat{a}^{+}$ are the photon annihilation and creation operators,
 $\zeta$   is the strength of the Stark shift, leading to the intensity-dependent transition frequency.
 The atomic two-level systems are described by the spin operators
 $\hat{s}^{z}=\frac{1}{2}\sigma^{z}$,
 $\hat{s}^{\pm}=\frac{1}{2}\big(\sigma^{x}\pm i\sigma^{y}\big)$, where
 $\sigma_{x,y,z}$ are Pauli operators. If the initial kinetic energy of the system is
 small $\frac{P^{2}}{2m}\ll
 d\sqrt{\frac{\hbar\omega_{0}n}{2\varepsilon_{0}V}}$, one can neglect the atomic motion and consider the standard TC model for uniform
 cavity $g(x)\approx\Omega_{0}$.
 Here $d$ is the atomic dipole moment, $V$ is the volume of the
 cavity, $\varepsilon_{0}$ is the electric constant and $n$ is the number of photons in the cavity.  In the opposite limit the coupling constant
 $g(x)$ depends on the position of the atom inside the cavity. Therefore, the motion of the atom has a strong impact on the inter level transitions leading to a  complex nonlinear dynamics.
 We show below that this nonlinearity and the complexity can be utilized for a spin orientation control.  We assume a standard form for the
 dependence $g(x)=\Omega_{0}\cos k_{f}x$,  where $k_{f}$   is the wave number in a lossless cavity of the Fabry-Perot type \cite{Prants}.
 Consequently, the operator of the coordinate $\hat{x}$  together with the field and the spin operators form a complete set of observables.
 In what follows we will derive the Heisenberg equations for each observable and investigate the model (\ref{eq:ham}) in the semi-classical approximation.
       We recall the standard commutation relations
       $\big[\hat{s}^{+};\hat{s}^{-}\big]=2\hat{s}^{z}$, $\big[\hat{s}^{z};\hat{s}^{\pm}\big]=\pm
       \hat{s}^{\pm}$, $\big[\hat{a};\hat{a}^{+}\big]=1$ and the
expressions for the operators in the interaction representation
$\hat{a}(t)=\hat{a}e^{-i\omega_{f}t}$,
$\hat{a}^{+}(t)=\hat{a}^{+}e^{i\omega_{f}t}$. From the Heisenberg
equations of motion
$\frac{d\hat{A}}{dt}=\frac{\partial\hat{A}}{\partial
t}+i\big[\hat{H};\hat{A}\big]$ and  eq. (\ref{eq:ham}) we
infer
\begin{eqnarray}\label{eq:motion}
\frac{dx}{d\tau}&=&\alpha p, \nonumber \\
\frac{dp}{d\tau}&=&-\sin x\cdot u,\nonumber \\
\frac{du}{d\tau}&=&-\big(\delta+\frac{\Delta}{2}+\Delta
s_{z}\big)v,\\
\frac{dv}{d\tau}&=&\big(\delta+\frac{\Delta}{2}+\Delta
s_{z}\big)u+16\cos x\cdot s_{z}\cdot (N-s^{2}+2s_{z}^{2}), \nonumber \\
\frac{ds_{z}}{d\tau}&=&-\cos x\cdot v. \nonumber
\end{eqnarray}
Here the semi-classical averaging procedure is used \cite{Prants}
$x=k_{f}\langle\hat{x}\rangle$,
$a_{y}=\langle\hat{a}_{y}\rangle$,$a_{x}=\langle\hat{a}_{x}\rangle$,$s_{x}=\langle\hat{s}_{x}\rangle$,
$s_{y}=\langle\hat{s}_{y}\rangle$,
$s_{z}=\langle\hat{s}_{z}\rangle$,
$p=\frac{\langle\hat{P}\rangle}{k_{f}}$ and the following notations
are introduced  $\alpha=\frac{k^{2}_{f}}{m\Omega_{0}}$,
$\delta=\frac{\omega_{f}-\omega_{0}}{\Omega_{0}}$,
$\frac{\zeta}{\Omega_{0}}=\Delta$,  $\tau=\Omega_{0}t$,
$u=4(a_{x}s_{x}+a_{y}s_{y})$, $v=4(a_{x}s_{y}-a_{y}s_{x})$,
$a^{2}_{x}+a^{2}_{y}-s^{2}_{z}=N=const$,
$s_{x}^{2}+s_{y}^{2}+s_{z}^{2}=const$. We are interested in  studying
the system described by eq. (\ref{eq:motion}) in different asymptotic limits. In particular, we focus on the
dynamically induced control of the spin orientation and on switching.
Our theoretical model can be realized easily in the experiment, e.g. by using Rydberg atoms and superconductive high-finesse cavities.
For instance one may consider  two-photon transitions between ${44}S_{1/2}\longleftrightarrow {43}S_{1/2}$ for rubidium atoms ${85}Rb$,
or ${40}S_{1/2}\longleftrightarrow  {39}S_{1/2}$ for cesium ${85}$ $Cs$. Such  transitions proceed through the intermediate  level
$39P_{3/2}$ or $43P_{3/2}$, and are of the type -two photons allowed and one photon forbidden- \cite{Metwally}.
 Depending on the principle quantum number $n$
the dynamical Stark shift is in the range $0<\zeta<100MHz$ \cite{Metwally}.
The cavity-atom coupling constant (which defines the time scale
of our problem) is $\Omega_{0}\sim 1-10 MHz $. Therefore, both limits $\Omega_{0}> \zeta$ and $\Omega_{0}<\zeta$ which
will be discussed below are realistic from the experimental point of view.

\section{Adiabatic solutions:  The Resonant case}
 In the semi-classical limit $N\gg s^{2}$
and within the regime of strong coupling $\alpha\ll1$
eqs. (\ref{eq:motion}) tend to the form
\begin{eqnarray}\label{eq:motion1}
\frac{dx}{d\tau}&=&\alpha p, \nonumber \\
\frac{dp}{d\tau}&=&-\sin x\cdot u,\nonumber \\
\frac{du}{d\tau}&=&-\big(g+\Delta
s_{z}\big)v,\\
\frac{dv}{d\tau}&=&\big(g+\Delta
s_{z}\big)u+16\cos x\cdot s_{z}\cdot N, \nonumber \\
\frac{ds_{z}}{d\tau}&=&-\cos x\cdot v. \nonumber
\end{eqnarray}
From these relations it follows  that the quantities

\begin{eqnarray} \label{eq:intmotion}
W=\frac{\alpha p^{2}}{2}-u\cos x+\frac{1}{2\Delta}(g+\Delta
s_{z})^{2},\;  \mbox{and} \\
u^{2}+v^{2}+16Ns^{2}_{z}=R^{2},~~~~~~~~~~~~~~~~~~\nonumber
\end{eqnarray}
are integrals of motion.  Here $g=\delta+\frac{\Delta}{2}$ is
 the de-tuning parameter.

The set of equations (\ref{eq:motion1}) is characterized by two time
scales. The
small parameter $\alpha\ll1$ sets the
time scale of the  slow center-of-mass motion of the atom; the fast time scale is determined
  by the inter-level
transitions.
The separation of time scales allows us to split the set of equations of motion  into
two parts
 \be
 \label{eq:xp} \frac{d^{2}x}{d\tau^{2}}+\alpha
u(\tau)\sin x=0. \ee
and
\be\label{eq:atom}\left\{\begin{array}{ll}
\frac{du}{d\tau}=-(g+\Delta s_{z})v, \\
\frac{dv}{d\tau}=(g+\Delta s_{z})u+16c N s_{z}, \\
\frac{ds_{z}}{d\tau}=-cv .\end{array} \right.
 \ee
 Due to the different time scales the slow parameter $c=\cos x$
 in eq.(\ref{eq:atom}) hardly varies on a reasonable time scales.
Since $\dot{x}\sim \alpha,$ a small change of the variable
$x(\tau)$ during the time interval $\tau<1/\alpha$
 can be neglected. The period of the inter-level transitions (see Eq.(12) below) is, however, shorter than $T\ll 1/\alpha$.
 Consequently, the system performs a large number of inter-level transitions with $ c=\cos(x(\tau)) \sim const $.
 Utilizing the integrals of motion (\ref{eq:intmotion}),
 one can derive a self-consistent equation for the spin operator
 $s_{z}(t)$:
 \be\label{eq:sz}
 \frac{ds_{z}}{d\tau}=-cv=-c\sqrt{R_{0}^{2}-\bigg[\frac{1}{2\Delta c}(g+\Delta s_{z})^{2}-\frac{W}{c}\bigg]^{2}-16 N s_{z}^{2}}
 .\ee
Here in the energy of the system (4) we neglect the adiabatic part
$\frac{\alpha p^{2}}{2}-u\cos(x)$. Consequently in the resonant
case, i.e. for $g=0$ we have: $W=\frac{\triangle}{2},~~~$
$\omega_{0}=\omega_{f}+\frac{\zeta}{2}$, and eq. (\ref{eq:sz}) can
be rewritten in the form
 \begin{eqnarray}\label{eq:sz1}
 \dot{s}_{z}&=&-cR_{0}\sqrt{(1-s_{z}^{2})(1-\kappa^{2}(1-s_{z}^{2}))},
 \nonumber\\
 \kappa &=&\frac{\Delta}{2cR_{0}},~~~R_{0}=4\sqrt{N}.
 \end{eqnarray}
 Consequently, the spin dynamics is described by the following solution
 \be \label{eq:solsz}
 s_{z}(\tau)=\left\{\begin{array}{l}
 cn(cR_{0}\tau,\kappa),~~~~\kappa<1,\\
dn(cR_{0}\tau\kappa,\frac{1}{\kappa}),~~~~\kappa>1.
 \end{array}\right.
 \ee
 Here, $cn(cR_{0}\tau,\kappa)$ and
 $dn(cR\tau\kappa,\frac{1}{\kappa})$ are periodic Jacobi elliptic functions \cite{Abramowitz}.
  For the other variables in eq. (6) we obtain (upon straightforward but laborious calculations) the expression
  \be\label{eq:u}
  u(\tau)=-\frac{\Delta}{2c}\left\{\begin{array}{l}
  sn^{2}(cR_{0}\tau,\kappa),~~~~\kappa<1,\\
   \frac{1}{\kappa^{2}}sn^{2}(cR_{0}\tau\kappa,\frac{1}{\kappa}),~~~~\kappa>1,
    \end{array}\right.
  \ee
  \be \label{eq:v}
  v(\tau)=\left\{\begin{array}{l}
  R_{0}sn(cR_{0}\tau;\kappa)dn(cR_{0}\tau;\kappa)~~~~~\kappa<1,\\
  \frac{R_{0}}{\kappa}sn(cR_{0}\tau\kappa,\frac{1}{\kappa})\sqrt{1-\kappa^{2}sn^{2}(cR_{0}\tau\kappa,\frac{1}{\kappa})}~~~~\kappa>1.
    \end{array}\right.
  \ee
From eq. (\ref{eq:solsz}) we conclude that, depending on the
parameters of the problem, the dynamics of the level populations
follow  different solutions. These solutions  are separated by the
special value $\kappa=1$ of the bifurcation parameter $\kappa$ that
signals  the presence of topologically distinct solutions.
In equation (\ref{eq:solsz}) the Jacobi elliptic functions are
periodic in the argument with the period \be
T=\left\{\begin{array}{l}\frac{4}{cR_{0}}K(\kappa),~~~~~\kappa<1,\\
\frac{2}{cR_{0}\kappa}K(\frac{1}{\kappa}),~~~~\kappa>1,
 \end{array}\right.
 \ee
 where  $K(\kappa)$ is  the complete elliptic integral of the first kind \cite{Abramowitz}.
  If $\kappa\rightarrow1$, the period becomes infinite because
  $K(\kappa)\rightarrow\ln(4/\sqrt{1-\kappa^{2}})$.
  The evolution in this special case is given by the non-oscillatory soliton
  solutions:
  \begin{eqnarray}
  s_{z}(\tau)&=&\frac{1}{\cosh(cR_{0}\tau)}, \nonumber \\
  v(\tau)&=&R_{0}\frac{\sinh^{2}(cR_{0}\tau)}{\cosh^{2}(cR_{0}\tau)},\\
  u(\tau)&=&-\frac{\Delta}{2c}\tanh^{2}(cR_{0}\tau). \nonumber
  \end{eqnarray}
  The existence of a bifurcation parameter indicates that the solutions separated by it,
  have different topological properties. The phase trajectories corresponding to the solution
  $dn(cR_{0}\tau\kappa,\frac{1}{\kappa})$ are open and they describe a rotational motion,
  while trajectories corresponding to $cn(cR_{0}\tau,\kappa)$
  are closed and they describe the oscillatory motion \cite{Zaslavsky}.
  Due to the three independent variables in eq. (\ref{eq:atom}) and the two integrals of motion (\ref{eq:intmotion}),
  the system, described by eq. (\ref{eq:atom}), is effectively one dimensional.
  Consequently, utilizing the integral of
  motion $R^{2}=R_{0}^{2}=16N$, $u(0)=0,~~v(0)=1,~~s_{z}(0)=1$
  one can easily reduce the system (eqs. (\ref{eq:atom})) to the effective one dimensional model
  \be \label{eq:Heff}
  H_{eff}=\frac{cv^{2}}{2}+\frac{\Delta^{2}}{8c}(1-s_{z}^{2})^{2}+\frac{c}{2}R_{0}^{2}s_{z}^{2}
  .\ee
 It is straightforward to conclude  that the solutions eqs.(\ref{eq:solsz})-(\ref{eq:v}) satisfy
   Hamilton's equations for taking the canonical pair of
  variables as $(s_{z},v)$
  \be
  \dot{s}_{z}=-H_{v}=-\frac{\partial H_{eff}}{\partial v},~~~~~~
  \dot{v}=H_{s_{z}}=\frac{\partial H_{eff}}{\partial s_{z}}
  .\ee
  Due to the nonlinearity, the model expressed  by eq. (\ref{eq:Heff}) shows a rich topological structure of
  phase-space trajectories. Topologically different types of phase trajectories are divided by  bifurcation points
  which can be identified by evaluating the following Poincare index
  \cite{Butenin}
  \be \label{eq:puancare}
  J=\frac{1}{2\pi}\oint d
  \big\{\tan^{-1}\big[-\frac{H_{s_{z}}}{H_{v}}\big]\big\}.
  \ee
  Here $\gamma$ is a contour around the equilibrium point $\dot{s}_{z}=0,
  \dot{v}=0$. From eqs. (\ref{eq:Heff}), and (\ref{eq:puancare})
  we conclude for eq. (\ref{eq:puancare}) in the linear approximation
  \be \label{eq:index}
  J\approx\frac{\cos^{2}(x)-\bigg(\frac{\zeta}{\Omega_{0}R_{0}}\bigg)^{2}}{\bigg|\cos^{2}(x)-
  \bigg(\frac{\zeta}{\Omega_{0}R_{0}}\bigg)^{2}\bigg|}. \ee
  The change of sign in $J$  from $J=1$ to $J=-1$ marks the transition from the stable
  equilibrium point to the unstable equilibrium saddle point, i.e.
  from  closed to  open phase trajectories.
  The bifurcation point is defined by the simple relation
  $x\sim\arccos\bigg(\frac{\zeta}{R_{0}\Omega_{0}}\bigg)$.
  Obviously,  the key issue is the relation between the position $x(t)$ of the atom inside the cavity
and the value of the dynamical shift of  the frequency $\zeta$.
For the determination of the transition time on a quantitative level we
need the exact solution of Eq. (\ref{eq:xp}).
 For further analysis of eqs. (\ref{eq:xp}), and (\ref{eq:atom}) we note that both solutions, given by eq. (\ref{eq:solsz}),
  contain the slow parameter
 $\kappa(\tau)=\big(\frac{\zeta}{2R_{0}\Omega_{0}\cos(x(\tau))}\big)^{1/2}$.
 If initially  $\frac{\zeta}{2R_{0}\Omega_{0}}<1$,
 $\cos(x(0))\approx1$, and $\kappa(0)<1$, even in this case, due to the adiabatic motion
 of the atom inside the cavity, the condition $\kappa(\tau)>1$ can be realized as well.
 This means that the motion of the atom inside a non-uniform cavity leads to
  a tunneling of the system, formally  through the presence of a separatrix.
 This leads to a qualitative change of the dynamics
 of $s_{z}(\tau)$, i.e. eq. (\ref{eq:solsz}). The explicit solution for
 $x(t)$ can be found from eq. (\ref{eq:xp}). Taking  eq. (\ref{eq:u}) into account and considering
 $c=\cos x $ as  an adiabatic parameter on the time scale
 $\tau\sim1/R_{0}$ from eq. (\ref{eq:xp}) we infer
 \be \label{eq:x}
 \frac{d^{2}x}{d\tau^{2}}+f(x)=f(x)\cos(2R_{0}\tau).
 \ee
Here $\varepsilon=\alpha\frac{\zeta}{\Omega_{0}}$,
$f(x)=-\varepsilon\tan x$ is the small parameter that controls the
time scale for the slow motion of the atom inside the cavity and
which is definitely larger than the time of the inter-level
transitions
$\sqrt{\frac{2\Omega_{0}}{\alpha\zeta}}>\frac{1}{R_{0}}$ (since
$N>1$, $\alpha<<1$). This means that during a slight change of the
center of mass position from its initial value $x(\tau=0)=0$,
 the system performs multiple inter level transitions described by the first solution of given by eqs. (\ref{eq:u}).\\
 For solving eqs. (\ref{eq:x}) we make use of the existence of a slow and  a fast time scale
 and  look for a solution of the  form
 \be \label{eq:solition}
 x(\tau)=x_{1}(\tau)+\mu x_{2}(\tau).
 \ee
The time scale for the slow variable $x_{1}(\tau)$ is governed by
 $1/\sqrt{\varepsilon}$. For the fast variable $x_{2}(\tau)$ we have
 $2R_{0}$, $2R_{0}\gg\sqrt{\varepsilon}$ and
 $\mu=\frac{\sqrt{\varepsilon}}{2R_{0}}$.
 With  $f(x_{1}+\mu x_{2})\approx f(x_{1})+\mu x_{2}\big(\frac{\partial f}{\partial
 x}\big)_{x_{1}}$ we infer
 from eq. (\ref{eq:solition}) that
 \be
 \ddot{x}_{1}+\mu \ddot{x}_{2}=-f(x_{1})-\mu x_{2}\bigg(\frac{\partial f}{\partial
 x}\bigg)_{x_{1}}+\bigg[f(x_{1})+\mu x_{2} \bigg(\frac{\partial f}{\partial
 x}\bigg)_{x_{1}}\bigg]\cos(2R_{0}\tau)
 .\ee
 After  averaging over the time interval $1/2R_{0}$ we find
 \be \label{eq:x1}
 \ddot{x}_{1}\approx-f(x_{1})+\bigg\langle\mu x_{2}\bigg(\frac{\partial f}{\partial
 x}\bigg)_{x_{1}}\cos(2R_{0}\tau)\bigg\rangle,
 \ee
 \be \label{eq:x2}
 \mu x_{2}\approx-\mu x_{2}+f(x_{1})\cos(2R_{0}\tau)
 .\ee
Considering that  $\mu \ddot{x}_{2}\sim \mu
(2R_{0})^{2}x_{2}\sim 2R_{0}\sqrt{\varepsilon}x_{2}>\mu
x_{2}\bigg(\frac{\partial f}{\partial x}\bigg)_{x_{1}}$ eq.
(\ref{eq:x2}).  The result is
\begin{eqnarray}\label{eq:integrate}
x_{2}(\tau)&=&-\bigg[\frac{f(x_{1})}{2R_{0}\sqrt{\varepsilon}}\bigg]\cos(2R_{0}\tau),
\nonumber \\
\ddot{x}_{1}(\tau)&+&f(x_{1})+\frac{1}{2}\frac{f(x_{1})}{4R_{0}^{2}}\bigg(\frac{\partial
f}{\partial x}\bigg)_{x_{1}}=0.
\end{eqnarray}
Finally, in the linear approximation $f(x)=-\varepsilon x$ we
conclude from eq. (\ref{eq:integrate}) that  \be \label{eq:xt}
x(\tau)=\sinh\bigg(\sqrt{\bigg(\frac{\zeta}{\alpha\Omega_{0}}-\frac{\zeta^{2}}{8R_{0}^{2}\Omega_{0}^{2}}\bigg)}\alpha\tau\bigg)
\bigg(1+\frac{\alpha\zeta}{4R_{0}^{2}\Omega_{0}}\cos(2R_{0}\tau)\bigg).
 \ee
Using eqs. (\ref{eq:sz1}), and (\ref{eq:xt}) we identify the time of
bifurcation via
 \be \label{eq:tau}
\tau_{b}\approx\sqrt{\frac{\Omega_{0}}{\alpha\zeta}}
\ln\bigg(\arccos\big(\frac{1}{2R_{0}}
\frac{\zeta}{\Omega_{0}}\big)\bigg),~~~k(\tau_{B})\sim 1.
 \ee
Eq. (\ref{eq:tau}) connects the values of the DSS parameter
$\zeta$ and the birfurcation time $\tau$. This means that by
observing the singularity in the spectrum of the Rabi oscillations
of the inter-level transitions one can measure the frequency shift
indirectly. Before proceeding with the non resonant case we
 summarize briefly the main results obtained for resonant case. The time
scale of the spatial motion is described by the parameter $\alpha
=\frac{k_{f}^{2}}{m\Omega_{0}}$. Since $\dot{x}\sim\alpha$ in the
limit of a strong coupling between the atom and the cavity
$\Omega_{0}>>\frac{k_{f}^{2}}{m}$, the motion of the atom inside
of the cavity is adiabatic. Therefore, the system performs a large
number of inter-level transitions before the slow adiabatic
coordinate changes significantly.  If the  resonant condition
$\zeta =2(\omega_{0}-\omega_{f})$ it met the inter-level
transitions are described by topologically different solutions
(cf. eq.(9)). The bifurcation parameter $k(x(\tau))$ depends on
the center-of-mass coordinate of the atom $x(\tau)$ (eqs.(24),
(25)). We argue that the existence of topologically different
solutions divided by a bifurcation point is the reason why the
spatial motion of the atom leads to singularities in the spectrum
of inter-level transitions.

\begin{figure}[t]
 \centering
  \includegraphics[width=8cm]{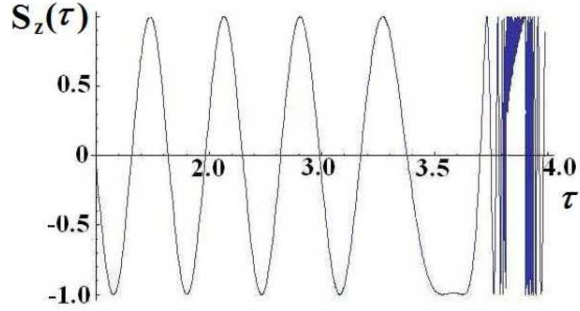}
  \caption{Color online. The time evolution of the atomic inversion, as descirbed by  Eqs.(\ref{eq:solsz},
  \ref{eq:xt}) for the following the parameters: $R_{0}=20$, $\alpha =5\cdot10^{-3}$, $\zeta/\Omega_{0}=2$.
  When  the parameter $\kappa(\tau)$ reaches the bifurcation values $\kappa(\tau_{b})=\bigg(\frac{\zeta}{2R_{0}\Omega_{0}\cos\big(x(\tau_{b})\big)}\bigg)^{1/2}\sim1$,
  $\tau_{b}\approx\sqrt{\frac{\Omega_{0}}{\alpha\zeta}}\ln\bigg(\arccos\big(\frac{1}{2R_{0}}\frac{\zeta}{\Omega_{0}}\big)\bigg)\sim4$ the inversion oscillations manifest a singularity.} \label{Fig:1}
\end{figure}
\section{Adiabatic solutions. Non resonant case }
\begin{figure}[t]
 \centering
  \includegraphics[width=8cm]{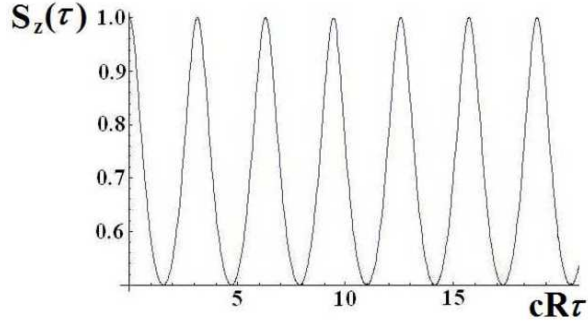}
  \caption{Switching of the spin projection $s_{z}(\tau)$ as induced by the motion of the atom inside the cavity.
  We have chosen  $\frac{g}{R_{0}c}\sim 1$.} \label{Fig:2}
\end{figure}

In the non-resonant case
$g=\frac{1}{\Omega_{0}}\bigg(\frac{\zeta}{2}+\omega_{f}-\omega_{0}\bigg)\neq0$,
the solution of the equation (\ref{eq:sz}) has the form \be
\label{eq:weiersrass}
s_{z}(\tau)=1-\frac{1}{2W(cR\tau;g_{2};g_{3})-A_{2}}
 .\ee
 Here $W(cR\tau;g_{2};g_{3})$ is the Weierstrass function
\cite{Abramowitz} and the following notations are used:
\begin{eqnarray}\label{eq:notations}
g_{2}&=&3A_{2}^{2}-2A_{1},~~~~~g_{3}=A_{1}A_{2}-A_{2}^{3}-\frac{1}{2}A_{0},
\nonumber \\
A_{2}&=&-\frac{1}{6}\bigg(1+\kappa^{2}\bigg(2+\frac{2g\Omega_{0}}{\zeta}\bigg)^{2}\bigg),~~~~~A_{1}=\frac{\kappa^{2}}{2}\bigg(2+\frac{2g\Omega_{0}}{\zeta}
\bigg),\\
A_{0}&=&-\kappa^{2},~~~~\kappa=\frac{\zeta}{2\Omega_{0}R_{0}\cos(x)}\nonumber
.\end{eqnarray}
For  $\kappa<<1$ i.e. $\zeta<<2\Omega_0R_{0}$,
$|A_{2}|\gg|A_{1}|$, $|A_{2}|\gg|A_{0}|$,
  $g_{2}=3A_{2}^{2}$ $g_{3}=-A_{2}^{3}$
and the roots of the equation
 \be \label{eq:Jacobi}
 4x^3-g_{2}x-g_{3}=0 , \ee
satisfy the condition
  $e_{1}=-A_{2}$,
  $e_{2}=e_{3}=\frac{A_{2}}{2}$,
  $A_{2}=\frac{1}{6}\bigg(1+\frac{g}{R_{0}c}\bigg)^2$.
Therefore we can use the following representation of the
Weierstrass function \cite{Abramowitz}:
  \be \label{eq:cot}
  W(\tau,g_{2},g_{3})=e_{1}+\frac{3}{2}e_{1}\cot^{2}\bigg(\sqrt{\frac{3}{2}e_{1}}\tau\bigg)
  .\ee
As a result finally from eq. (\ref{eq:weiersrass}) we obtain
 \be \label{eq:cot1}
  s_{z}(\tau)=1-\frac{1}{3e_{1}+\frac{3}{2}e_{1}\cot^{2}\bigg(\sqrt{\frac{3}{2}e_{1}}cR\tau\bigg)}
  .\ee
From this expression  we see that the motion of the atom leads to a switching of the spin orientation between the values
 $s_{z}(\tau)=1$ and  $s_{z}(\tau)=1-\frac{1}{3e_{1}}$. The time interval for the switching is set by the simple relation
$T=\frac{\pi}{2cR_{0}\sqrt{6e_{1}}}$. An illustration of this
switching is shown by the simulations presented in
Fig.\ref{Fig:2}.
This dynamically induced switching can be utilized for the
manipulation of the spin orientation. A further  important issue
is the principle difference between the results obtained for the
non-resonant case ($\zeta\neq\omega_{f}-\omega_{0} $) from those
results corresponding to the resonant case
($\zeta=2(\omega_{f}-\omega_{0}) $).  In contrast to the resonant
case, for the non-resonant situation the spatial motion of the
atom leads to a switching of the spin orientation. However, due to
the absence of a bifurcation and topologically different
solutions, the singularities in the spectrum of inter-level
transition are absent.

\section{Dynamics for  small DSS and minimal chaos}
\begin{figure}[t]
 \centering
  \includegraphics[width=8cm]{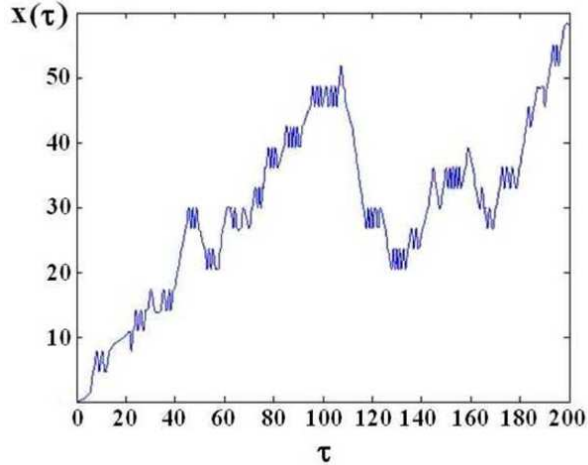}
  \caption{Color online. The chaotic motion of the atom inside the cavity in the regime of a weak coupling.
  The numerical  integration of eqs. (\ref{eq:motion}) is performed for the following values of
  the parameters $\delta=-0.5$, $\alpha=0.5$, $N=50$, $s=1$, $x(0)=0$, $p(0)=1$,$u(0)=0$, $v(0)=0$, $s_{z}(0)=1$.
 As evident from the figure,  the motion in the nonadiabatic case is chaotic and resembles  a diffusion process.} \label{Fig:3}

\end{figure}

\begin{figure}[t]
 \centering
  \includegraphics[width=8cm]{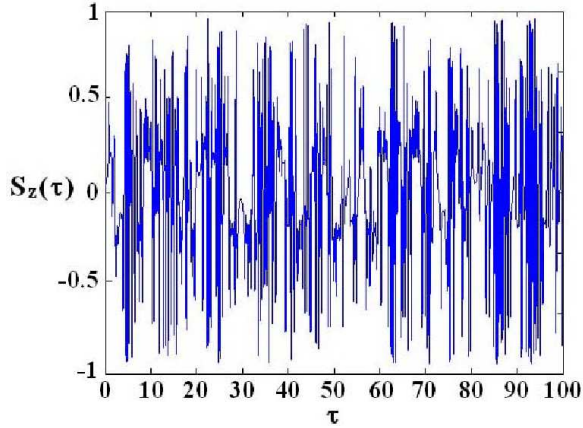}
  \caption{Color online. The dynamics of the atomic inversion as deduced from the numerical integration of
 eqs.(\ref{eq:motion}) for the following parameters values:
  $\delta=-0.5$, $\alpha=0.5$, $N=50$, $s=1$, $x(0)=0$, $p(0)=1$,$u(0)=0$, $v(0)=0$, $s_{z}(0)=1$.
  Obviously the inter-level transitions are of a chaotic nature.  } \label{Fig:4}

\end{figure}

\begin{figure}[t]
 \centering
  \includegraphics[width=8cm]{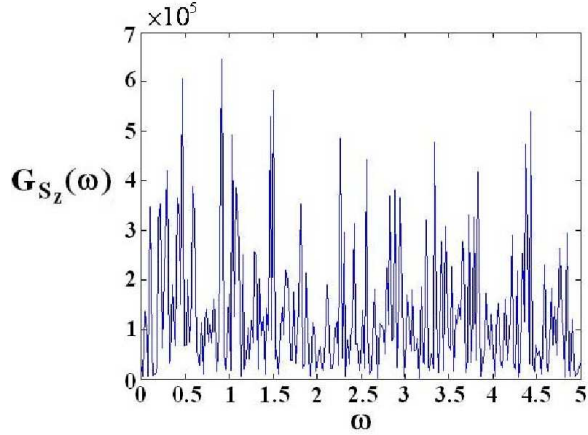}
  \caption{The correlation function for the atomic inversion
 as delivered by the numerical integration of eqs.(\ref{eq:motion}) for  $\delta=-0.5$, $\alpha=0.5$, $N=50$, $s=1$, $x(0)=0$, $p(0)=1$,$u(0)=0$, $v(0)=0$, $s_{z}(0)=1$.
  The finite  width of the correlation function signifies the dynamical stochasticity.} \label{Fig:5}

\end{figure}

If the detuning between the radiation field and the spin splitting is larger
than DDS, i.e. for $\delta=\frac{\omega_{f}-\omega_{0}}{\Omega_{0}}\gg\frac{\zeta}{\Omega_{0}}$,
the  equations of motion reduce to
\be
 \label{eq:xp1}1) \frac{d^{2}x}{d\tau^{2}}+\alpha
u(\tau)\sin x=0, \ee \be \label{eq:atom1}
2)\left\{\begin{array}{ll}
\frac{du}{d\tau}=-\delta v,  \\
\frac{dv}{d\tau}=\delta u+16c N s_{z}, \\
\frac{ds_{z}}{d\tau}=-\cos x v . \end{array}
 \right.
 \ee
 Since $\alpha\ll1,$  the adiabatic approximation is still valid and the eqs.
 (\ref{eq:atom1}) is analytically integrable. Introducing
 $u(\tau)=X(\tau)$, $v(\tau)=Y(\tau)$, $Z(\tau)=R_{0}s_{z}$, $C_{0}=-R_{0}\cos x$
 the solution can be written in the compact matrix form
 \be \label{eq:sol1}
 \left(\begin{array}{ll}
 X(\tau)\\Y(\tau)\\Z(\tau)\end{array}\right)=M(\tau) \left(\begin{array}{ll}
 X(0)\\Y(0)\\Z(0)\end{array}\right)
 \ee
 \be \label{eq:sol2}
 M(\tau)=\left(\begin{array}{cc}\frac{C_{0}^{2}}{\Omega_{N}^{2}}+\frac{\delta^{2}}{\Omega_{N}^{2}}\cos\Omega_{N}\tau
 ~~~-\frac{\delta}{\Omega_{N}}\sin\Omega_{N}\tau ~~~
 \frac{C_{0}\delta}{\Omega_{N}^{2}}(1-\cos\Omega_{N}\tau)\\
\frac{\delta}{\Omega_{N}}\sin\Omega_{N}\tau ~~~ \cos\Omega_{N}\tau
~~~
-\frac{C_{0}}{\Omega_{N}}\sin\Omega_{N}\tau \\
\frac{C_{0}\delta}{\Omega_{N}^{2}}(1-\cos\Omega_{N}\tau) ~~~
\frac{C_{0}}{\Omega_{N}}\sin\Omega_{N}\tau ~~~
\frac{\delta^{2}}{\Omega_{N}^{2}}+\frac{C_{0}^{2}}{\Omega_{N}^{2}}\cos\Omega_{N}\tau
\end{array}\right)
 \ee
 where $\Omega_{N}^{2}=C_{0}^{2}+\delta^2=\delta^2+R_{0}^{2}\cos^{2}
 x$. For the particular initial conditions $u(0)=v(0)=0$,
 $s_{z}(0)=1$, the solution (\ref{eq:sol2}) simplifies and reduces to the compact
 form:
 \begin{eqnarray}\label{eq:sol3}
 u(\tau)&=&-\frac{16N\delta}{\Omega_{N}^{2}}\big(1-\cos(\Omega_{N}\tau)\big),~~~~
 v(\tau)=-\frac{C_{0}}{\Omega_{N}}\sin\Omega_{N}\tau,\nonumber\\
 s_{z}&=&\frac{\delta^{2}}{\Omega_{N}^{2}}+\frac{C_{0}^{2}}{\Omega_{N}^{2}}\cos\Omega_{N}\tau.
 \end{eqnarray}
 Using (\ref{eq:sol3}), equation (\ref{eq:xp1}) can be rewritten as
 \be \label{eq:xp2}
 \frac{d^2x}{d\tau^2}-\omega^2(1-\cos\Omega_{N}\tau)\sin x=0.
 \ee
  Here  $\omega=\frac{4R_{0}}{\Omega_{N}}\sqrt{\alpha|\delta|}$.
 Equation (\ref{eq:xp2})  corresponds effectively to the perturbed universal Hamiltonian
  \begin{eqnarray}\label{eq:pert}
  H&=&H_{0}+V(\tau),\nonumber \\
  H_{0}&=&\frac{1}{2}\dot{x}^2-\omega^2\cos x,\\
V(\tau)&=&\frac{\omega^2}{2}\big[\cos(x+\Omega_{N}\tau)+\cos(x-\Omega_{N}\tau)\big].\nonumber
    \end{eqnarray}
    This model shows a behaviour known as minimal chaos, as discussed in details in  Ref.\cite{Zaslavsky}.
    The width of the stochastic layer formed near to the separatrix due to the time dependent
    perturbation $V(\tau)$ can be estimated via the following
    expression:
    \be
    \Delta
    E=\int\limits_{-\infty}^{+\infty}\{H_{0};V\}=-\int\limits_{-\infty}^{+\infty}\dot{x}
    \frac{\omega^2}{2}\big(\sin(x+\Omega_{N}\tau)+\sin(x-\Omega_{N}\tau)\big)d\tau=
-\omega^2\int\limits_{\infty}^{\infty}\dot{x}\sin(x-\Omega_{N}\tau)
    .\ee
    Using the separatrix solutions for the unperturbed part of Hamiltonian
    $H_{0}$ (\ref{eq:pert}), \cite{Zaslavsky}
    \begin{eqnarray}\label{eq:separatrix}
    x_{s}&=&4\arctan\exp\big[\pm\omega(\tau-\tau_{0})\big],\nonumber\\
    \dot{x}_{s}&=&\pm\frac{2\omega}{\cosh\big[\omega(\tau-\tau_{0})\big]},
    \end{eqnarray}
we obtain \be\label{eq:layer} \delta
E_{s}\leq\frac{4\pi\Omega_{N}^{3}}{\omega}\frac{\exp(\pi\Omega_{N}/2\omega)}{\sinh\big(\frac{\pi\Omega_{N}}{\omega}\big)}.\ee
Only in the energy interval located near the separatrix
$|E-E_{s}|<\delta E_{s}$, the motion is irregular. However, in the
non-adiabatic case, i.e. if $\alpha=\frac{k_{f}^2}{m\Omega_{0}}$ is
not small anymore, the dynamics is irregular in the whole phase
space, as illustrated in Figs.(\ref{Fig:3}, \ref{Fig:4}, and \ref{Fig:5})

To confirm  the existence of chaos, we examine the width of
Fourier  transform of the correlation function
$G_{s_{z}}(\tau')=\langle s_{z}(\tau+\tau')|s_{z}(\tau) \rangle $,
$G_{s_{z}}=\int\limits_{-\infty}^{+\infty}d\tau
G_{s_{z}}(\tau)\exp[i\omega\tau]=\frac{\tau_{c}}{1+\omega^{2}\tau_{c}^{2}}$.
With
$\langle\ldots\rangle=\lim\limits_{T\rightarrow\infty}\frac{1}{T}\int\limits_{0}^{+\infty}(\ldots)dt$
we mean  the averaging with respect to time, while $\tau_{c}$
being the correlation time. The result of numerical calculations
is presented in Fig.\ref{Fig:5}. The finite width of the Fourier
transform signifies the emergence of chaos. The numerical results
presented in Figs.\ref{Fig:3}-\ref{Fig:5} confirm that in the
limit of weak coupling $\alpha=\frac{k_{f}^{2}}{m\Omega_{0}}\sim
1$  the dynamics turns chaotic.  In contrast, in the limit of  a
strong
coupling 
chaos appears only in a small area near to the separatrix, as
quantified by eq.(\ref{eq:layer}). For nonzero DSS the dynamics is
integrable in both cases: for the resonant case
$\omega_{0}=\omega_{f}+\frac{\zeta}{2}$ (\ref{eq:solsz}) and for
the non-resonant case $\omega_{0}\neq\omega_{f}+\frac{\zeta}{2}$
(\ref{eq:weiersrass}).

Summarizing this part of the work, we found  that when the
detuning between the  radiation field and the spin splitting is
larger than DSS $\omega_{f}-\omega_{0}>\zeta$ two different types
of dynamics are realized:
 1) In the regime of a strong coupling between the atom and
the cavity $\alpha=\frac{k_{f}^{2}}{m\Omega_{0}}\ll 1$ the
dynamics is regular in the whole phase space, except for a narrow
energy gap near to the classical separatrix $|E-E_{s}|<\delta
E_{s}$ (cf. Eqs. (39), Eq.(40)). 2) In the regime of a weak
coupling $\alpha\sim1$ almost the whole phase space shows chaotic
dynamics. Due to the impact of the spatial motion on the spin
dynamics, the spectrum of the inter-level transitions is chaotic
as well (see Fig.4).

If the mean photon number in the cavity is not large, then the
semi-classical approximation is not justified and  the problem
should be considered quantum-mechanically. A relevant  question in
this case is the quantum correlation within the compound quantum
systems, i.e. between the atom and the radiation field. The
entanglement between the atom and the field is usually quantified
in terms of purity \cite{Uleysky,Argonov}. As entanglement is a
specific quantum form of correlation it exhibits  a number of
essential differences to classical correlations.
These issues  will be the subject of the next section.\\

\section{Entanglement between the field and the atom:  Quantum mechanical consideration}
As was mentioned above, if the mean photon number inside  the cavity is
not a large number, then a  semi-classical
treatment is not viable  and one should resort to quantum mechanics
\cite{Uleysky,Argonov}.
Having said that,  the translational degrees of freedom   can still be treated classically;
 the inter-level transitions are described quantum mechanically, however.
  Thus,  the center-of-mass coordinate of the atom $x(t)$
may still be described by the  classical Hamilton's equations of motion
\be \dot{x}=\alpha p,~~~~ \dot{p}=-\langle\hat{u}\rangle \sin
x.\ee
The  quantum
mechanical average should be performed then, i.e.   \be
\langle\hat{u}\rangle=\bigg\langle\frac{1}{\sqrt{N}}\big(e^{i\delta
t}s^{+}a^{2}+e^{-i\delta t}s^{-}(a^+)^{2}\big)\bigg\rangle . \ee
 The average is taken with respect to  the wave functions that solve for the Schr\" odinger equation
 \be\label{eq:schrodinger}
 i\frac{d|\psi(t)\rangle}{dt}=\hat{H}_{int}|\psi(t)\rangle.
 \ee
 Here
 \be\label{eq:interaction}
 \hat{H}_{int}=\zeta\hat{a}^{+}\hat{a}s_{z}+g(x)\big(e^{i\delta
 t}s^{+}\hat{a}^{2}+e^{-i\delta
 t}s^{-}(\hat{a}^{+})^{2}\big),~~~~\delta=\omega_{0}-2\omega_{f}.
 \ee

 Following the standard procedure of Ref. \cite{Schleich}, we  seek  solutions of the equation (\ref{eq:schrodinger}) in the form
 \be\label{eq:wave}
 |\psi(t)\rangle=\sum\limits_{n}\bigg[C_{e,n}(t)|e,n\rangle+C_{g,n+2}(t)|g,n+2\rangle\bigg].
 \ee
 Substituting eq. (\ref{eq:wave}) into eq. (\ref{eq:schrodinger}) we find a set of equations for determining the expansion
 coefficients
 \begin{eqnarray}\label{eq:C}
 i\dot{C}_{e,n}&=&\zeta\frac{n}{2}C_{e,n}+g(x)\sqrt{(n+1)(n+2)}e^{-i\delta
 t}C_{g,n+2}, \nonumber \\
i\dot{C}_{g,n+2}&=&-\zeta\frac{n+2}{2}C_{g,n+2}+g(x)\sqrt{(n+1)(n+2)}e^{i\delta
 t}C_{e,n}.
 \end{eqnarray}
  In order to derive analytical results for the degree of entanglement we will study
  the eqs. (\ref{eq:C}) in two different limits: For a strong atom-cavity coupling
  $\alpha=\frac{k_{f}^{2}}{m\Omega_{0}}\ll1$,  the quantity $x(t)$   is an adiabatic variable on the time scale
   set by the inverse Rabi frequencies.
 In this case system, as described by eqs. (\ref{eq:C}), is exactly solvable and the solutions are given by the following
  expressions
  \begin{eqnarray}\label{eq:solC}
  C_{e,n}(t)&=&e^{i\frac{(\delta-\zeta)t}{2}}\Big\{-\frac{i
  b(n)}{\lambda_{n}}\sin(\lambda_{n}t)C_{g,n+2}(0)+\bigg(\cos(\lambda_{n}t)-\frac{i
  (\delta+\zeta(n+1))\sin(\lambda_{n}t)}{2\lambda_{n}}\bigg)C_{e,n}(0)\Big\},\nonumber\\
    C_{g,n+2}(t)&=&e^{-i\frac{(\delta+\zeta)t}{2}}\Big\{-\frac{i
  b(n)}{\lambda_{n}}\sin(\lambda_{n}t)C_{e,n}(0)+\bigg(\cos(\lambda_{n}t)+\frac{i
  (\delta+\zeta(n+1))\sin(\lambda_{n}t)}{2\lambda_{n}}\bigg)C_{g,n+2}(0)\Big\},\nonumber\\
  b(n)&=&g(x)\sqrt{(n+1)(n+2)},~~~\lambda_{n}=\frac{\sqrt{(\delta+\zeta(n+1))^{2}+4b^2(n)}}{2}.
  \end{eqnarray}
  \begin{figure}[t]
 \centering
  \includegraphics[width=8cm]{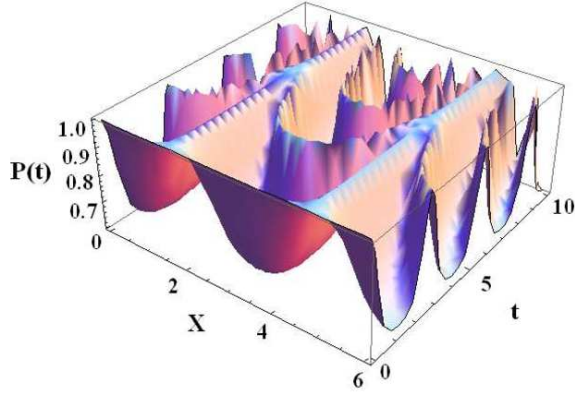}
  \caption{Color online. The purity $P(t)$ as a function of time $t$ and of the center-of-mass
   position of the atom, as dictated by eq. Eq.(51).  The following   parameters have been chosen
    $\Omega_{0}=1$ ,$\bar{n}=1$, $\zeta=1$. } \label{Fig:6}
\end{figure}
  Using these solutions and  after tracing out the field's degrees of freedom
  one can  introduce  the reduced density matrix for the atomic
  subsystem as
  \be\label{eq:density}
  \rho^{a}=\left(\begin{array}{c}\rho_{11}^{a}~~~\rho_{12}^{a}\\
  \rho_{21}^{a}~~~\rho_{22}^{a}\end{array}\right)
  ,\ee
  where
  \begin{eqnarray}\label{eq:rho}
  \rho_{11}^{a}&=&\sum\limits_{n}\bigg|C_{e,n}(t)\bigg|^{2},~~~
  \rho_{12}^{a}=\sum\limits_{n}C_{e,n}(t)C^{*}_{g,n+2},\nonumber\\
  \rho_{21}^{a}&=&\sum\limits_{n}C_{g,n+2}^{*}(t)C_{e,n},~~~
  \rho_{22}^{a}=\sum\limits_{n}\bigg|C_{g,n+2}(t)\bigg|^{2},\\
  C_{e,n}(0)&=&c_{e}W_{n},~~~C_{g,n+2}(0)=c_{g}W_{n+2},c_{e}=1,c_{g}=0.\nonumber
    \end{eqnarray}
    Here $W_{n}^{2}=\frac{\bar{n}^n}{n!}e^{-\bar{n}}$ is the distribution function of the field coherent states \cite{Schleich}.
     Substituting eq.(\ref{eq:solC}) in eq.(\ref{eq:rho}) we conclude  the following explicit form for the  elements of  the reduced density matrix (\ref{eq:density}) $(\omega_{0}=2\omega_{f}+\zeta,n\gg1)$:

          \begin{eqnarray}\label{eq:rhomatrix1}
     \rho_{11}^{a}&=&\frac{1}{2}\bigg(1+e^{-\bar{n}(1-\cos(2t\lambda))}\cdot\cos(\bar{n}\sin(2t\lambda)\bigg)+\nonumber\\
     &+&\frac{1}{2}\frac{\zeta^2}{4\lambda^2}\bigg(1-e^{-\bar{n}(1-\cos(2t\lambda))}\cdot\cos(\bar{n}\sin(2t\lambda)\bigg),\nonumber\\
      \rho_{12}^{a}&=&-\frac{i g(x)}{2\lambda}e^{-i\zeta t}e^{-\bar{n}(1-\cos(2t\lambda))}\sin(\bar{n}\sin(2t\lambda)+\\
      &+&\frac{1}{2}e^{-i\zeta
      t}\cdot\frac{g(x)\zeta}{2\lambda^2}\bigg(1-e^{-\bar{n}(1-\cos(2t\lambda))}\cdot\cos(\bar{n}\sin(2t\lambda)\bigg),\nonumber\\
       \rho_{21}^{a}&=&\frac{i g(x)}{2\lambda}e^{i\zeta t}e^{-\bar{n}(1-\cos(2t\lambda))}\sin(\bar{n}\sin(2t\lambda)+\nonumber\\
      &+&\frac{1}{2}e^{i\zeta
      t}\cdot\frac{g(x)\zeta}{2\lambda^2}\bigg(1-e^{-\bar{n}(\cos(2t\lambda)-1)}\cdot\cos(\bar{n}\sin(2t\lambda)\bigg),\nonumber\\
        \rho_{22}^{a}&=&\frac{1}{2}\frac{g^{2}(x)}{\lambda^2}\bigg(1-e^{-\bar{n}(1-\cos(2t\lambda))}\cdot\cos(\bar{n}\sin(2t\lambda)\bigg),\nonumber\\
        \lambda^2&=&\bigg(\frac{\zeta}{2}\bigg)^{2}+g^2(x).\nonumber
     \end{eqnarray}

     The interaction between the atom and the radiation field is  described by the nonseparable wave function (45), i.e. by an
      entangled state \cite{Schleich}. The entanglement between the atom and the  field has a different meaning
     from the usual definition of the entanglement between the atomic states. The field is the subsystem with a large number of degrees of freedom and is  usually  prepared in a coherent state. Therefore, the state of the field is not influenced by the atom-field coupling interaction. For quantifying the entanglement between the atom and the field we should average and trace out the field's states on the cost of a partial loss of coherence. To make this  point clear,  let us consider the simplest protocol of a quantum measurement \cite{Schleich}.
     The  probability $W(t,|\psi_{a}(t)\rangle, |\psi_{field}(t)\rangle)$ that both subsystems are
     in the particular states $|\psi_{a}(t)\rangle $ and $|\psi_{field}(t)\rangle $ is defined via the following relation:
     \be
      W(t,|\psi_{a}(t)\rangle, |\psi_{field}(t)\rangle)=|\langle\psi_{field}\mid\langle\psi_{a}\mid\psi(t)\rangle|^{2},
      \ee

     \be
     \mid\psi(t)\rangle =\sum\limits_{n}\sum\limits_{j=e,g}\psi_{jn}(t)\mid j,n \rangle .
     \ee
     Here
     \be
     \psi_{en}(t)=C_{en}(t),~~~\psi_{gn}(t)=C_{g,n+2}(t).
     \ee
     As simple quantum measurement one may perform on the system is that, one observes   the atomic state for an arbitrary field
     state.
       As we already have mentioned above, the field is prepared in a coherent state  \cite{Schleich}.
    The entanglement between the atom and the field (while statistically averaging over the field states) is quantified in terms of the purity $P$ \cite{Bruss}
     \be
     P\big(t,\mid\psi_{a}\rangle \big)=\sum\limits_{n=0}\limits^{\infty}\mid\sum\limits_{j=e,g}\psi_{jn}(t)\langle\psi_{a}\mid j,n \rangle\mid^{2}.
     \ee
Comparing eq. (54) with  eq. (51) we conclude on a partial loss of coherence,
since some interference terms are omitted in eq.(54).
 Nevertheless, the atomic and
the field subsystems are still entangled and the density operator
of the total system cannot be written as a direct product of
density operators each corresponding to
 the atom and the field subsystems
     Using eqs. (49, 53 and 54) the purity of the quantum state is expressible in terms
     of the reduced density matrix in the standard form \cite{Bruss}, meaning that
     \begin{eqnarray}
     &&P\big(t,\mid\psi_{a}\rangle \big)=P(t)=Tr\bigg(\big(\rho^a\big)^{2}\bigg)=\big(\rho^{a}_{11}\big)^{2}+\big(\rho_{22}^{a}\big)^{2}+2\rho_{12}^{a}\rho_{21}^{a}=\nonumber\\
     &&=1-\frac{g^{2}(x)}{\lambda^{2}}\big\{1-\exp \big[-2\bar{n}\big(1-\cos(2t\lambda)\big)\big]\big\},\\
     &&0<P(t)<1. \nonumber
     \end{eqnarray}
From eq.(55) we readily deduce  that the degree of coherence degrades   with the decrease of DSS.
 The purity as a function of time
and of  the adiabatic coordinate is plotted in Fig.6.
The purity is distributed non-uniformly, since the  Rabi frequencies are coordinate dependent (\ref{eq:solC}).
 In the semi-classical limit, when the mean photon number in the cavity is large
$\bar{n}\gg 1$, the expression given by eq. (55) simplifies and for the time- and the coordinate-averaged purity we obtain  $P=\bigg(1-\frac{2\Omega_{0}^{2}}{\zeta^{2}+4\Omega_{0}^{2}}\bigg)$.
We see that the ratio of the cavity-atom coupling constant and the DSS
 $\Omega_{0}/\zeta$ is still the determining factor in our problem.
 From the expression above, the maximum of the purity is $P_{max}=1$ for the case $\Omega_{0}/\zeta\ll 1$ and $P_{min}=\frac{1}{2}$ for $\Omega_{0}/\zeta\gg 1$,  respectively.
In the opposite case  corresponding to the weak atom-cavity coupling
      with $\alpha =\frac{k_{f}^{2}}{m\Omega_{0}}\sim1$, the analytical solution of the system (\ref{eq:C}) can be found in the special case of a resonant driving
     $\delta=\omega_{0}-2\omega_{f}=0$
     \begin{eqnarray}\label{eq:weak}
     C_{e,n}&=&\exp\bigg(-\frac{in\zeta t}{2}\bigg)\bigg(C_{1}Q[n\omega(t)]+C_{2}Q^{*}[n\omega(t)]\bigg),\nonumber
     \\
C_{g,n+2}&=&\exp\bigg(\frac{i(n+1)\zeta
t}{2}\bigg)\bigg(C_{1}Q[n\omega(t)]-C_{2}Q^{*}[n\omega(t)]\bigg),
     \end{eqnarray}
     where
     \be\label{eq:C12}
     C_{1}=\frac{C_{e,n}(0)+C_{g,n+2}(0)}{2},~~~C_{2}=\frac{C_{e,n}(0)-C_{g,n+2}(0)}{2},
     \ee
     and
     \be\label{eq:Q}
     Q[n\omega(t)]=\exp\Big[in\int\limits_{0}^{t}\omega(t')dt'\Big],~~~Q^{-1}[\omega]=Q^{*}[\omega],~~~\omega(\tau)=\cos(x(\tau)).
     \ee
     From eqs.(\ref{eq:weak}, \ref{eq:C12}) we infer for the reduced density matrix and its purity the following forms
\begin{eqnarray}
     &&P(t)=Tr\bigg(\big(\rho^a\big)^{2}\bigg)=\big(\rho^{a}_{11}\big)^{2}+\big(\rho_{22}^{a}\big)^{2}+2\rho_{12}^{a}\rho_{21}^{a}\\
     &&0<P(t)<1. \nonumber
     \end{eqnarray}
     The explicit forms of the density matrix elements are
     \begin{eqnarray}\label{eq:entrho}
     \rho_{11}^{a}&=&\sum\limits_{n}\bigg[C_{1}^2+C_{2}^2+C_{1}C_{2}\big(Q^{*2}\big[n\omega(t)\big]+Q^{2}\big[n\omega(t)\big]\big)\bigg],\nonumber\\
      \rho_{12}^{a}&=&\sum\limits_{n}\bigg[C_{1}^2-C_{2}^2+C_{1}C_{2}\big(Q^{*2}\big[n\omega(t)\big]-Q^{2}\big[n\omega(t)\big]\big)\bigg], \nonumber \\
      \rho_{21}^{a}&=&\sum\limits_{n}\bigg[C_{1}^2-C_{2}^2+C_{1}C_{2}\big(Q^{2}\big[n\omega(t)\big]-Q^{*2}\big[n\omega(t)\big]\big)\bigg],\\
 \rho_{22}^{a}&=&\sum\limits_{n}\bigg[C_{1}^2+C_{2}^2-C_{1}C_{2}\big(Q^{*2}\big[n\omega(t)\big]+Q^{2}\big[n\omega(t)\big]\big)\bigg].\nonumber
     \end{eqnarray}
\begin{figure}[t]
 \centering
  \includegraphics[width=8cm]{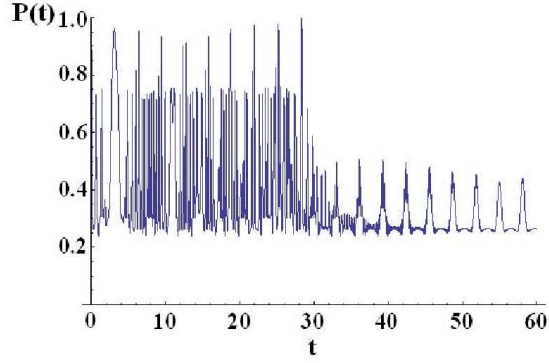}
  \caption{Color online. The purity $P(t)$ as a function of time $t$, as described by  Eqs.(55, 57, 58).  The following parameter are chosen: $\alpha=10^{-2}$ $\bar{n}=1$, $\zeta=0.2$, $\Omega_{0}=0.1$.} \label{Fig:7}
\end{figure}

\begin{figure}[t]
 \centering
  \includegraphics[width=10cm]{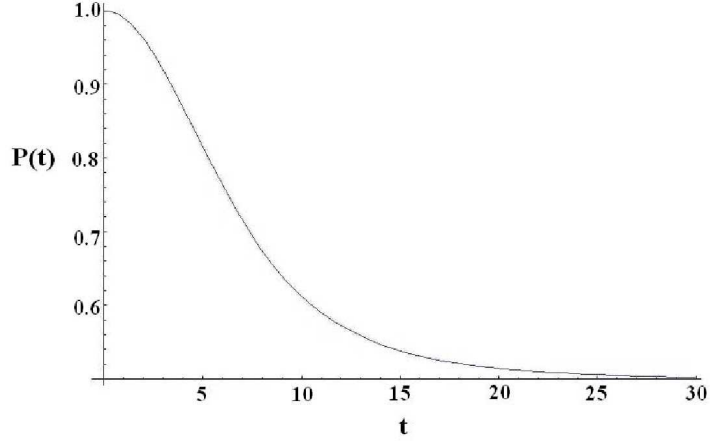}
  \caption{The purity $P(t)$ as a function of the time $t$.  The results follow from  Eqs.(55, 57, 61). In the calculations the
   following parameters  are chosen
   $\alpha=0.5$ $\bar{n}=1$, $\zeta=0.2$, $\Omega_{0}=0.1$, $\alpha_{0}=1$.} \label{Fig:8}
\end{figure}
Considering the initial conditions (\ref{eq:C12}), and tracing
over the field states in eq. (\ref{eq:entrho}),  we obtain for
matrix elements
\begin{eqnarray}\label{eq:rhomatrix2}
     \rho_{11}^{a}&=&\frac{1}{2}+\frac{1}{4}e^{-\bar{n}}\bigg(e^{\bar{n}Q[2\omega[t]]}+e^{\bar{n}Q^{*}[2\omega[t]]}\bigg),\nonumber\\
      \rho_{22}^{a}&=&\frac{1}{2}-\frac{1}{4}e^{-\bar{n}}\bigg(e^{\bar{n}Q[2\omega[t]]}+e^{\bar{n}Q^{*}[2\omega[t]]}\bigg),\\
       \rho_{12}^{a}&=&\frac{1}{4}e^{-\zeta i}\bigg(e^{\bar{n}Q^{*}[2\omega(t)]\exp(-it\zeta)}-e^{\bar{n}Q[2\omega(t)]\exp(-it\zeta)}\bigg)e^{-\bar{n}},\nonumber\\
        \rho_{21}^{a}&=&\frac{1}{4}e^{\zeta i}\bigg(e^{\bar{n}Q[2\omega(t)]\exp(it\zeta)}-e^{\bar{n}Q^{*}[2\omega(t)]\exp(it\zeta)}\bigg)e^{-\bar{n}}.\nonumber
     \end{eqnarray}
The time dependence of the matrix elements of the reduced density matrix, as given by eqs.
(\ref{eq:rhomatrix1}, \ref{eq:rhomatrix2}), is governed  by
the exponential factors (\ref{eq:Q}). We evaluate $Q[2\omega(t)]$ for
two different regimes. In the regime of a regular motion, using
solution (\ref{eq:xt}) we simply have
 \be\label{eq:Qt}
Q\big[2\omega(t)\big]\approx\exp\Big[2i\sqrt{\frac{\Omega_{0}}{\alpha\zeta}}
\big[Ci(e^{t\sqrt{\frac{\alpha\zeta}{\Omega_{0}}}})-Ci(1)\big]\Big]
,\ee
  where $Ci(\ldots)$ is the cosine integral function \cite{Abramowitz}.
   In this case  the purity  exhibits fast oscillations, as evidenced by Fig.7.
From Fig.7 we conclude  that after a lap of time $t=t_{c}\approx 50$, the oscillation frequency decreases.
The reason for this behavior can be traced back to the time dependency  of the parameters given by eq. (58) (i.e.,  $Q\big[ 2\omega (t) \big]$):
 For $t>t_{c}$, $Q\big[ 2\omega(t)\big]\approx \exp \big[-2i\sqrt{\frac{\Omega_{0}}{\alpha \zeta}} Ci(1)\big]=const$
 and the diagonal elements of the density matrix are constant in time
\be
\rho^{a}_{11}=\frac{1}{2}+\frac{1}{2}\exp(-\bar{n})\cos\big(\bar{n}\exp\big[2\sqrt{\frac{\Omega_{0}}{\alpha\zeta}}Ci(1)\big]\big),
\ee
\be
\rho^{a}_{22}=\frac{1}{2}-\frac{1}{2}\exp(-\bar{n})\cos\big(\bar{n}\exp\big[2\sqrt{\frac{\Omega_{0}}{\alpha\zeta}}Ci(1)\big]\big).
\ee
 Hence,  $\big( \rho_{11}^{a}\big)^{2}+\big( \rho_{22}^{a}\big)^{2}= const$ and the temporal dependence
of the purity is governed by the off-diagonal term $2\rho_{12}^{a}\rho_{21}^{a}$.
In the semi-classical limit $\bar{n}\gg 1$ the purity turns constant $P=1/2$ and is independent of the values of the frequency shift.
   In the regime when the motion is chaotic, we use a more advanced technique for the evaluation of
  $Q[\omega(t)]$. Namely, because of the random character of the atomic motion, the exponent, given by eq. (\ref{eq:Q}),
    should be considered as a functional of the random function $\omega(t)$.
   Therefore, we should carry out a statistical average with respect to all possible realizations of the random
   parameters. The mean value of the functional $\langle Q[\omega(t)]
   \rangle$ can be calculated by evaluating the following integral
   \begin{eqnarray}\label{eq:continual}
   \langle
   Q[\omega(t)]\rangle=\exp\Big[i\int\limits_{0}^{t}\omega(t')dt'\Big]=\lim\limits_{N\rightarrow\infty\atop\Delta t_{k}\rightarrow0}\int d\omega_{N}\ldots
   d\omega_{1}\exp\Big[i\sum\limits_{k=1}^{n}\omega_{k}\Delta
   t_{k}\Big]P_{N}[\omega], \nonumber\\
   \Delta t_{k}=t^{k}-t^{k-1}, ~~~ t^{0}=0,~~~t^{N}=1.
   \end{eqnarray}
   Here, the multidimensional normal distribution function is given via  the  expression
   \be\label{eq:distribution}
   P_{N}[\omega]=(2\pi)^{-N}\int d\lambda_{1}\ldots
   d\lambda_{N}\exp\Big[-i\sum_{k}\lambda_{k}\omega_{k}\big]\exp\Big[-\frac{1}{2}\sum\limits_{k,k'}C_{kk'}\lambda_{k}\lambda_{k'}\Big],
   \ee
     where $\lambda_{k}$ are distribution parameters and $C_{kk'}$
     is the covariation matrix \cite{Feller}.
      Substituting eq. (\ref{eq:distribution}) in eq. (\ref{eq:continual}) and performing the integration we obtain
\be \label{eq:Q1} \langle
Q[\omega(t)]\rangle\approx\exp\Big[-\frac{t}{2}\sqrt{\frac{\pi}{\alpha_{0}}}Erf\bigg[t\sqrt{\alpha_{0}}\bigg]\Big]
,\ee where $Erf[\ldots]$ is the error function, and $1/\alpha_{0}$
is the width of the correlation function \\
$c(\tau)=\langle\omega(t+\tau)\omega(t)\rangle$. Taking eqs.
(\ref{eq:rhomatrix2}-\ref{eq:Q1}) into account one can evaluate
the time dependence of the  purity in both the regular and the
chaotic cases. The result of the numerical calculations is
presented in Fig.(\ref{Fig:8}). As we see, after a fast decay, the
purity  stabilizes around  the value $P(t)\sim 0.5$. The
corresponding decay rate is determined by the correlation width
$1/\alpha_{0}$ of the random parameter $x(t)$.

\section{Conclusions}
In the present paper, our aim was to analyze the dynamics of the
TC atom in a non-uniform cavity, in order to uncover the
consequences of the DSS  for the weak and the strong cavity-field
couplings. We found that in the regime of a strong
coupling $\alpha=\frac{k_{f}^{2}}{m\Omega_{0}}\ll 1$, the motion of
the atom inside the cavity is adiabatic. For the case   when an
exact resonance is reached $\zeta=2(\omega_{0}-\omega_{f})$, then
depending on the values of the center of mass coordinate of the
atom $x(t)$, the level populations are described by topologically
distinct solutions separated by a bifurcation parameter.
The motion of the atom inside the non-uniform cavity
leads to a tunneling of the system through the separatrix and as a
consequence to singularities in the inter-level transition
spectrum $S_{z}(t)$.  The bifurcation point is
identified by a simple analytical expression, given by eq.(25). A key
factor  is the relationship  between DSS and the atom-cavity
coupling constant $\zeta/\Omega_{0}$. Therefore, by observing the
singularity in the spectrum of the Rabi oscillations of the
inter-level transitions $S_{z}(t)$ one may infer the frequency
shift in an indirect way. Far from the resonance
$\zeta>>2(\omega_{0}-\omega_{f})$, the spatial motion of the atom
inside the cavity leads to a  switching of the spin projection
between the values  $S_{z}=1$ and $S_{z}=1/2$. The period of switching
depends on the values of DSS:
$T=\frac{\pi}{2(R_{0}cos(x)+g)},~g=
\frac{\zeta}{\Omega_{0}}+\frac{\omega_{f}-\omega_{0}}{\Omega_{0}}$.
In the regime of a  weak coupling
$\alpha=\frac{k_{f}^{2}}{m\Omega_{0}}\sim1$, the motion of the
atom inside the cavity becomes chaotic (see Fig. 3-5). We find
that, in the adiabatic case, the ratio of the cavity-atom coupling
constant and the DSS $\Omega_{0}/\zeta$ are still the determining  factors.
Namely, in the semi-classical limit, when the mean photon number
in the cavity is large $\bar{n}>>1$, the time and the
coordinate-averaged purity is
$P=\bigg(1-\frac{2\Omega_{0}^{2}}{\zeta^{2}+4\Omega_{0}^{2}}\bigg)$.
Therefore, the maximum of the purity  $P_{max}=1$ is achieved for
the case $\Omega_{0}/\zeta<<1$ and $P_{min}=\frac{1}{2}$
corresponds to the case  $\Omega_{0}/\zeta>>1$.

\textbf{Acknowledgments} The financial support by the Deutsche
Forschungsgemeinschaft (DFG) through SFB 762, the HGSFP (grant
number GSC 129/1), the Heidelberg Center for Quantum Dynamics,
grant No. KO-2235/3 and STCU grant No 5053 is gratefully
acknowledged.

\end{document}